\documentclass[12pt,a4paper]{article}

\usepackage[left=1.2in, right=1.2in]{geometry}

\usepackage{amssymb}
\usepackage{amsfonts}
\usepackage{amsmath}
\usepackage{graphicx}
\usepackage{subcaption}
\usepackage[font={small,bf}]{caption}
\usepackage{multirow}
\usepackage{algorithm}
\usepackage{algpseudocode}
\usepackage{bm}
\usepackage{varwidth}

 \usepackage[disable]{todonotes} 

\usepackage{placeins}

\algtext*{EndFor}
\algtext*{EndIf}

\DeclareMathOperator*{\argmax}{arg\,max}
\DeclareMathOperator*{\argmin}{arg\,min}

\begin{document}

\title{An Experimental Evaluation of Regret-Based Econometrics}

\author{
Noam Nisan\thanks{Rachel \& Selim Benin School of Computer Science \& Engineering and Federmann Center for the Study of Rationality, The Hebrew University of Jerusalem, Israel; and Microsoft Research.  Supported by ISF grant 1435/14 administered by the Israeli Academy of Sciences and by
Israel-USA Bi-national Science Foundation (BSF) grant number 2014389.}
\and
Gali Noti\thanks{Rachel \& Selim Benin School of Computer Science \& Engineering and Federmann Center for the Study of Rationality, The Hebrew University of Jerusalem, Israel; and Microsoft Research.  Supported by the Adams Fellowship Program of the Israel Academy of Sciences and Humanities. }
}

\maketitle
\begin{abstract}
Using data obtained in a controlled ad-auction experiment that we ran, we evaluate the regret-based approach to econometrics that was recently suggested by Nekipelov, Syrgkanis, and Tardos (EC 2015).
We found that despite the weak regret-based assumptions, the results 
were (at least) as accurate as those obtained using classic equilibrium-based assumptions. 
En route we studied to what extent humans actually minimize regret in our ad auction, and 
found a significant difference between the ``high types'' (players with a high valuation) who indeed rationally minimized regret and
the ``low types'' who significantly overbid.  
We suggest that correcting for these biases 
and adjusting the regret-based econometric method 
may improve the accuracy of
estimated values.
\end{abstract}


\section{Introduction}

The field of econometrics combines observational data with specific modeling assumptions in order 
to estimate parameters of interest.  It goes beyond mere statistical analysis in that it assumes specific
models of how the parameters of interest relate to the type of data observed.  
On the one hand, these assumptions
provide power to econometric analysis by allowing it to estimate parameters that do not 
directly appear in the data,
but, on the other hand, the correctness of the whole estimation strongly depends on the
correctness of the utilized model, i.e., on 
the extent to which the situation at hand indeed conforms
to the theoretical model.

In game-like scenarios, typical models assume that players are at an equilibrium.  This is of course
not a trivial assumption as we know that humans are not fully rational and certainly
do not always ``find'' an equilibrium point (see, e.g., \cite{Kagel1995, KL12} and the references therein).  Beyond the usual 
human lack of rationality, equilibrium assumptions are especially hard to justify
in complex scenarios such as those found in electronic auctions.  This may be due to a variety of reasons
such as computational hardness, dependence on private information or the prior,
non-intuitiveness of the situation, 
or repeated aspects of the game.  

A recent paper \cite{Nekipelov2015} suggested using a much
weaker assumption than the equilibrium assumption.  This was demonstrated in the specific 
context of ad auctions (of the type that are also known as ``sponsored-search auctions''), but can be applied to every repeated-game scenario.  The weaker assumption
that they promoted was that players {\em minimize regret} in the specific sense used in the 
regret-minimization literature \cite{BM2007} (sometimes known as ``Hannan consistent'').  
This notion
assumes that players manage to achieve at least as much utility as they could have gotten from 
playing any {\em fixed} action repeatedly.   This makes minimal
assumptions about the players' learning and rationalizing ability.  Certainly, if the players reach a
Nash equilibrium, they must all be minimizing their regrets, but the regret-minimization
assumption is strictly weaker, and, for example, holds even if the players 
reach any equilibrium from the much wider families of correlated or even
coarse equilibria.

In \cite{Nekipelov2015}, ad-auction data from Microsoft was analyzed under the regret-minimization assumption with the goal
of estimating advertisers' values that are not known to the search engine (Microsoft), by using the 
bids that are known to the search engine.  Classic econometric methods were applied to this
task in \cite{Varian2007} and \cite{Athey2010}, 
and the average value estimates that were obtained in \cite{Nekipelov2015} seemed to be more or less 
in line with those from 
the two other methods (\cite{Varian2007} and \cite{Athey2010}). 
However, since their data set lacked the real values
of the advertisers
it is not clear how well each of these three econometric methods did, and, in particular, it
is not clear how good are the estimates that were obtained based on the weak regret-minimization
assumption.

It turned out that we had previously performed an experiment that 
yields exactly the type of data that can allow evaluation of the success of econometric methods in an ad-auction context \cite{NNY2014}. 
In our experiment (described in Section \ref{sec:exp}), human players participated in a simulation of a stream of ad auctions. Unlike in field data, in our controlled experiment we assigned valuations to the players, 
and so we can directly compare the value estimates of the econometric methods with these real (assigned) values. 

In the present paper we use this experimental data to evaluate the performance of the regret-based method for estimating players' values, 
in comparison with the classic equilibrium-based estimation methods. 
We start by asking to what extent humans succeed in minimizing their regret in the repeated auction, and find that the answer depends on the ``type'' (private value) of the player (Section \ref{sec:regret}): while higher-type players indeed rationally minimize 
their regret, the lower-type players remain with high levels of regret. In Section \ref{sec:method} we specify our implementation of the regret-based estimation method, and in Section \ref{sec:eval} we evaluate this method in comparison with the classic equilibrium-based methods. 
We find that the regret-based method manages to perform 
at least as well as the other methods, even in the truthful VCG auction where the equilibrium prediction is particularly strong. 
That is, our findings suggest that the weaker assumptions and general approach of the regret-based method are sufficient for this estimation task. In addition, we find that while all methods perform reasonably well for the higher-type players, they result in high errors for the lower-type players. In Section \ref{sec:possible_improvements} we suggest several approaches to improve the accuracy of the estimates, using the regret-based 
econometrics  
and the understanding of the type-based bias 
that we identified.


\section{Our Experiment}\label{sec:exp}

\begin{figure}[t]
\centerline{\includegraphics[scale=0.55]{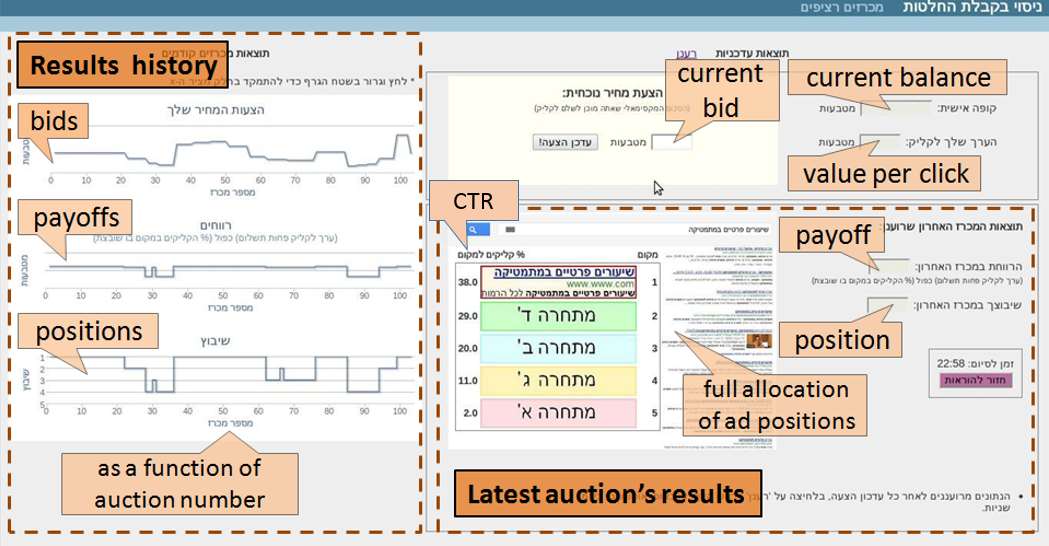}}
\caption{Game interface. \label{fig:screenshot} }
\end{figure}

We performed a controlled experiment where human subjects were asked to participate in a simulation
of ad auctions, similar to those held by search engines like Google or Microsoft.  This experiment was described in
\cite{NNY2014}, which contains all the details as well as the results.  
In the experiment, we recruited participants in groups of five.  
In each instance of the experiment, 
the five participants simulated the roles of advertisers and had to 
compete in a stream of ad auctions that lasted 25 minutes. 
We used a flexible 
auction experimentation software 
that we developed that 
enabled us to control the auction details as well 
as the players' knowledge and values. 
The auctions were conducted continuously, one auction per 
second, 
to a total of 1500 auctions within the 25-minute game. 
The participants could modify their bids at any time, 
and each auction was performed with the current settings of the bids. 
Each player was assigned a ``type'' at random, 
which was
his private ``valuation,'' i.e., the monetary value that he obtained from each user 
who clicked on his ad (we used 21, 27, 33, 39, 45 ``coins''). 
Each ad auction sold 
five ad positions with varying (commonly known) Click Through Rates (CTR) (we used $\bm{\alpha}=(2\%, 11\%, 20\%, 29\%, 38\%)$), 
which were displayed in a decreasing order of CTRs, so that the position on the top of
the page received the highest CTR. 
Every time an advertiser with a valuation $v$ won a position 
with CTR $\alpha_k$, he got an income of $\alpha_k \cdot v$ from
that auction.  This income was added to his balance
and the appropriate payment according to the auction rule was deducted from his balance. 
The players were given a graphical user interface in which they could 
modify their bids as often as 
they wished, and follow the results of the auctions so far. 
Figure \ref{fig:screenshot} shows a screen shot of the user interface. 

The experiment had a two-way (2x2) between-participant design;  
thus there were four experimental conditions. 
The two factors were:

\begin{enumerate}
\item {\bf Payment Rule (the Auction Mechanism):} We compared the (theoretically appealing) VCG payment rule with the (commonly used) GSP payment rule.
Both VCG and GSP auctions make the same allocation of positions -- by decreasing order of bids -- but their payment rule is different. Unlike GSP, the VCG is truthful; i.e., in every VCG auction it is a dominant strategy for every player to bid his true value (see \cite{EOS2007,Varian2007}).
\item {\bf Valuation Knowledge:} While the starting point of analyzing behavior in auctions is the ``valuation'' of the bidder, it is questionable to
what extent users are explicitly aware of this valuation.  We compared the case where bidders were directly given their valuation (given value, GV), 
and were explained its significance,
and the case where bidders were not directly given the valuation, but rather only see their payoffs -- information from 
which the valuation may be deduced, but could alternatively be directly used to guide the bidding (deduced value, DV).
\end{enumerate}

There were a total of 24 experimental sessions, 6 sessions for each of the
4 experimental conditions (thus there were 12 sessions for each factor). 
The groups (of five players each) were randomly assigned to the 
four experimental conditions, giving a total of $n$ = 120 participants. For further details regarding the experimental setup see \cite{NNY2014}.

\begin{figure}[t]
\begin{subfigure}{.47\linewidth}
  \includegraphics[scale=0.5]{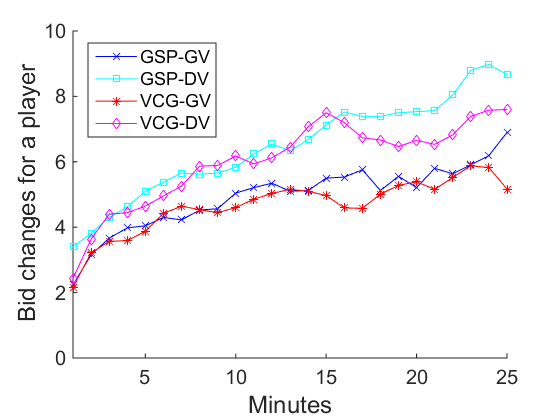}
  \caption{Bid modifications frequency: The average number of bid changes per minute made by a player, in each of the experimental conditions over time.}  
  \label{fig:bidActWithTime}
\end{subfigure}
\hspace{0.2cm}
\begin{subfigure}{.47\linewidth} 
  \includegraphics[scale=0.5]{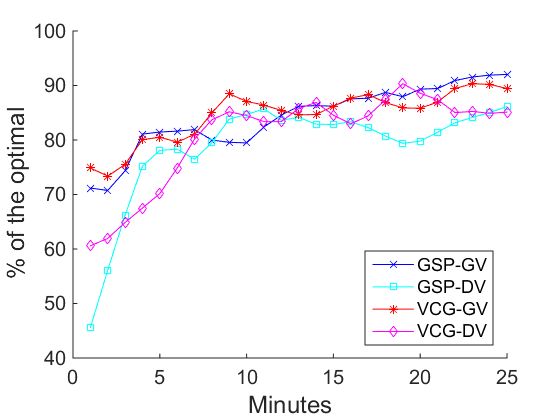}
  \caption{Social welfare: The average social welfare achieved in each of the experimental conditions over time, on a linear scale from worst to best outcomes.}
  \label{fig:swByCond}
\end{subfigure}%
\caption{Previous results.$^{\ref{footnoteSmooth}}$}
\label{fig:prev-results}
\end{figure}

\addtocounter{footnote}{1}\footnotetext{
Figures \ref{fig:bidActWithTime} and \ref{fig:swByCond} are taken from \cite{NNY2014}.\label{footnoteSmooth}}

Out of the various results of this experiment, let us mention two
that have particular relevance to the question at hand regarding the assumptions of
the econometric model.  The first result is that players in no way seemed to converge to an equilibrium.
In fact, players kept modifying their bids throughout the 25 minutes, and the frequency of
bid modification increased over time.  Figure \ref{fig:bidActWithTime} shows the average activity level 
(i.e., frequency of bid modification) as a function of time
in the four experimental conditions.  This was true in all auction formats, even
when using the VCG auction rule with explicitly given valuations for which truthful bidding
is a dominant strategy and so we would have expected truthful bidding as a strong
and stable equilibrium prediction.  

On the other hand, despite any lack of convergence to equilibrium,
it seems that the auction was able to quickly achieve
close to the ``correct''  (social-welfare maximizing) allocation of the slots,
attaining over 80-90\% of the optimal level of welfare\footnote{
Our count here is within the possible range of allocations, such that a random
allocation would get 50\% of the welfare.  In absolute terms, around 90-95\% of the welfare
was achieved.  See \cite{NNY2014} for details.} toward the end of the session (see Figure \ref{fig:swByCond}), 
as well as extracting revenue to the auctioneer that matches and even exceeds the theoretically expected
revenue.  This suggests that the different auction formats
were able to extract the value information from the users and utilize it, despite 
not having reached an equilibrium.  This would thus seem to indicate that it should be possible to 
deduce the (hidden) value information from the (visible) auction bids, even though
an equilibrium assumption does not seem to hold.

\section{To What Extent Do Humans Minimize Regret?}\label{sec:regret}

We start by asking to what extent does the assumption that players minimize regret in the repeated game hold in our experiment with human players? 
I.e. we compare the actual utility achieved by each one of our participants to the optimal utility he could have achieved had he played an optimal fixed bid in a sequence of auctions. 
More formally, let $\bm{b}^t$ denote the vector of actual bids played by the five players at time $t$, 
and denote by $U_i(b,\bm{b}^t_{-i}|v_i)$ the utility of player $i$ whose value is $v_i$ by bidding $b$ at time $t$.
Thus, player $i$'s {\em actual utility} in a sequence of auctions $(\bm{b}^t)_t$ is $Actual_i((\bm{b}^t)_t|v_i)=\sum_tU_i(b^t_i,\bm{b}^t_{-i}|v_i)$, where $b_i^t$ is the actual bid played by bidder $i$ at time $t$.
The {\em optimal utility} that player $i$ could have obtained by playing a fixed bid repeatedly 
in all auctions $(\bm{b}^t)_t$ is $Opt_i((\bm{b}^t)_t|v_i) = max_b\sum_t U_i(b,\bm{b}^t_{-i}|v_i)$. 
The {\em regret} of player $i$ whose value is $v_i$ in a sequence of auctions $(\bm{b}^t)_t$ is defined to be the difference between his optimal and actual utilities in these auctions, i.e., 
$$Regret_i((\bm{b}^t)_t|v_i)= Opt_i((\bm{b}^t)_t|v_i) - Actual_i((\bm{b}^t)_t|v_i)$$

For the VCG 
auction formats, the optimal fixed bid is certainly the dominant strategy 
of bidding the true value, but for the GSP auction
formats the optimal bid depends on the bids of the other players and
is only evident in hindsight (and thus could not even theoretically
be known to our bidders in real time.) 

We define the {\em regret of a set $S$ of players} just additively over the players in $S$. 
That is, $Actual_S=\sum_{i\in S}Actual_i$, and $Opt_S=\sum_{i\in S}Opt_i$. Then, the regret of $S$ at a sequence of auctions $(\bm{b}^t)_t$ is
$Regret_S((\bm{b}^t)_t) = Opt_S((\bm{b}^t)_t) - Actual_S((\bm{b}^t)_t) = \sum_{i\in S}Regret_i((\bm{b}^t)_t|v_i)$. 
Since the level of regrets depends on the magnitude of the utilities, we present the regret levels as percentages of the corresponding optimal outcomes. Specifically, we define the {\em relative regret} of $S$ at a sequence of auctions $(\bm{b}^t)_t$ by
$$RelativeRegret_S((\bm{b}^t)_t) = \frac{Regret_S((\bm{b}^t)_t)}{Opt_S((\bm{b}^t)_t)}$$

\begin{figure}[t]
\begin{subfigure}{.5\linewidth} 
  \includegraphics[scale=0.53]{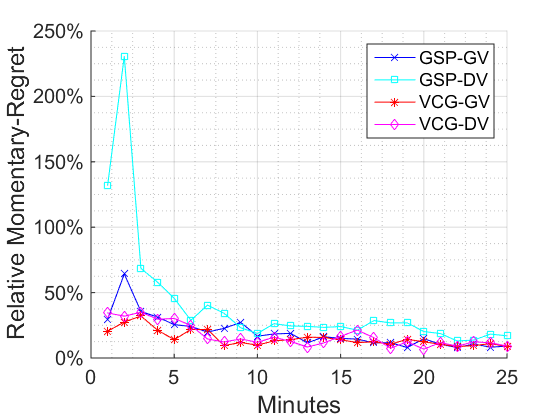} 
  \caption{By experimental conditions}
  \label{fig:momentary_regret_cond}
\end{subfigure}%
\begin{subfigure}{.5\linewidth} 
  \includegraphics[scale=0.53]{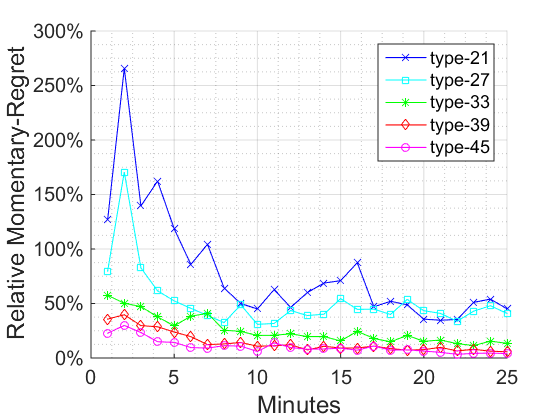}
  \caption{By player types}
  \label{fig:momentary_regret_type}
\end{subfigure}
\caption{Relative momentary regret over time. The momentary regret is the gap, computed separately for each minute, between the utility achieved by the players and the utility that would have been achieved had they played the fixed strategy that was optimal for that minute.
}
\label{fig:momentary_regret}
\end{figure}

First, in order for any regret minimization to be possible, we would need to see some learning by the bidders as time progresses, and indeed that is what we find.  
Figures \ref{fig:momentary_regret_cond} and \ref{fig:momentary_regret_type} show the
relative ``momentary regret'' of the players over time by experimental conditions and by player types, respectively, 
where the momentary regret is the regret computed separately in each minute of the experiment.
Consistent with the regret-minimization assumption, we find that the momentary regret decreased over time
in all conditions and types. 
Specifically,  for each of the four experimental conditions, the average momentary regret in the first third of the game is significantly higher than in the last third\footnote{When considering momentary regret, the first and last thirds of the game refer to the first and last 8 minutes, respectively. The results are robust to other choices of partitioning.} 
(N=6 sessions, Wilcoxon paired two-sided signed rank test, $p<0.05$ except for VCG-GV for which $p=0.06$) and the same for each of the five types of players (N=24, $p<0.001$ for each type except for 33 for which $p<0.03$). 
The decrease was faster at the beginning when players acquired experience in the game. 
This seems to be consistent with the suggestion raised by \cite{Nekipelov2015}, that higher levels of regret may indicate that bidders are in their initial learning phase. 
The regret toward the end of the game reached low levels of around 15\% regret for the different conditions, but, as can be seen in Figure \ref{fig:momentary_regret_type}, remained substantial for the lower-type players.

\begin{figure}
\begin{subfigure}{.48\linewidth}
  \includegraphics[scale=0.53]{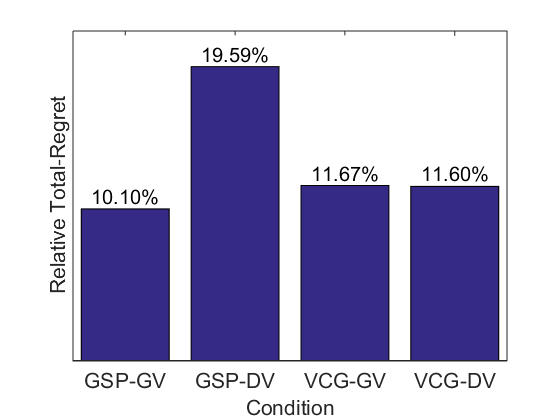}
  \caption{By experimental conditions}  
  \label{fig:total_regret_cond}
\end{subfigure}
\begin{subfigure}{.48\linewidth} 
  \includegraphics[scale=0.53]{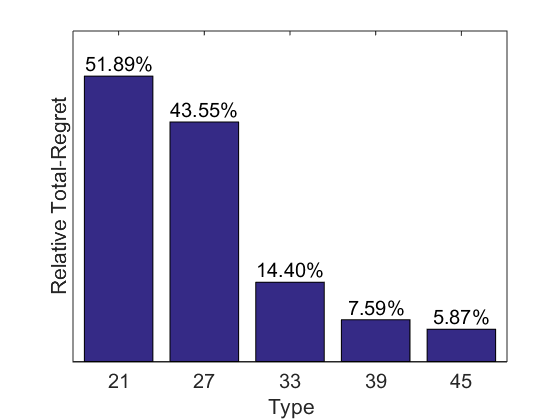}
  \caption{By player types}
  \label{fig:total_regret_type}
\end{subfigure}%
\caption{Relative total regret of the bidders according to experimental conditions and according to types of players. The total regret is computed over the second half of the auctions game (750 auctions), and is presented as a percentage of the corresponding optimal outcome.}
\label{fig:total_regret}
\end{figure}

At this point we can look at the overall regret achieved by our players, and answer the question to what extent is it true that bidders in ad auctions minimize regret, at least after the initial learning phase.  
We compute the ``total regret'' of a bidder based on the second half of the game (750 auctions), 
where regret levels stabilized, indicating that players had completed the initial learning phase and were experienced enough in the game.\footnote{
Qualitatively our findings are robust to modifications of the definition of the initial learning phase.} 
Looking according to the experimental conditions, the total regret seems reasonably low: a 10-20\% loss (Figure \ref{fig:total_regret_cond}). 
The total regret in GSP-DV was higher than in the other three conditions, as could be expected, since this condition 
combines the two difficult settings -- GSP and DV -- where 
bidders needed to learn both their opponents' behavior and their own values. 
Yet, the difference was significant at the 5\% level only in comparison with the GSP-GV condition. 


However, a different story is revealed when examining bidders' regret by their types: we found that the type of a player, i.e., the value assigned to him at random and used to determine his payoffs in all auctions, also had a significant effect on his regret-minimization performance. 
Figure \ref{fig:total_regret_type} shows the relative total regret for the five types of players. It can be seen that the lower the type is, the higher the player's total regret: the highest-type players achieved low levels of regret -- a less than 6\% loss -- while the lowest-type players remained with very high levels of regret -- with a more than 50\% loss of their optimal outcome.\footnote{Notice that the high percentage of regret of low-type players is not so high (though still higher) in absolute additive terms, since they tend to win 
the low CTR slots anyway. } 
Specifically, there is a significant negative correlation between the player's type and his total regret ($N=120$, $r=-0.68$, $p<0.001$). These differences between types were consistent throughout the game, as can be seen in Figure \ref{fig:momentary_regret_type}.
Therefore, high levels of regret may indicate not only that a player is in his initial learning phase, as suggested by \cite{Nekipelov2015}, but also that he might be of a lower type relative to his opponents.

Thus, we see that players who were assigned (by chance) with ``poor'' types turn out to play significantly
less rationally than players who were lucky to get ``rich'' types.  
While we were surprised at first to see such a
significant gap, on second thought this behavior seems quite intuitive: 
when the low-value players play rationally they tend to ``lose'' in the auction, i.e., win the low CTR slots.  This is quite frustrating and so they keep trying to ``win,'' but to no avail since they ``should'' be losing the
auction according to their true value.  
This gap in rational play 
is consistent with other irrational behaviors reported in \cite{NNY2014} -- overbidding and high frequency of bid changes -- that were also correlated with the player types. 
It may be interesting to relate this to other settings where it was found that
``the poor act irrationally'' (see, e.g., \cite{Barr2012, Bertrand2004, Katz1994}). 
However, our controlled setting proves that this ``irrational'' behavior cannot be explained by
any characteristics of the poor themselves (e.g., lower education), 
but rather only being poor (in an inferiority) relative to the others affected players' rationality.
These findings may be explained in terms of ``auction fever'' \cite{KMM2005, HYA2004}, 
 as a form of an ``illusion of control'' bias \cite{Langer1975}, or 
in terms of a stronger bias toward ``winning.''


Finally, we should also note here that the regret is positive: in principle it is possible to achieve negative regret (in the GSP auction) using time-varying bids. Yet not a single one of the 60 players who participated in our GSP auctions managed to achieve negative total regret.  I.e., none of our players managed to utilize the {\em dynamics} in the repeated game to their advantage and to play better than the fixed-bid benchmark.


\section{Regret-Based Econometrics}\label{sec:method}

Before we proceed to the evaluation results, let us formally specify our implementation of 
the regret-based method, suggested by \cite{Nekipelov2015}, for estimating bidders' valuations from their bids.
For this estimation task, we now assume that we observe only the bids played by the bidders and the fixed CTRs.

\begin{figure*}[t]
\begin{center}
\includegraphics[scale=0.59]{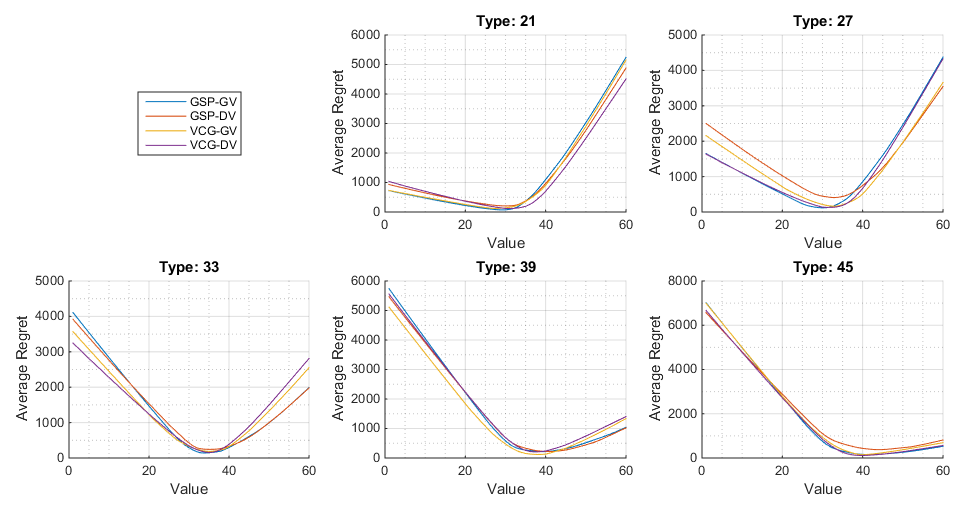}
\end{center}
\caption{Average total regret as a function of value according to player types and experimental conditions. The value that gives the minimal total regret for a player is given as the estimate for the player's value. \label{fig:regret-vs-v}}
\end{figure*}

The regret-based estimation method is based on the assumption that players use learning strategies that minimize their regret in the repeated game. That is, they use bidding strategies by which, over time, their utilities are not much worse than the optimal utilities they could have achieved by playing the best fixed bid in hindsight. 
Thus, the method sets the value estimate for a bidder, in a sequence of auctions, to be the value that minimizes his regret in these auctions.
Again, this assumption is weaker than the standard approach in econometrics, which relies on the assumption that the observed game reaches a static equilibrium, and it is general for different games (and, in particular, for different auction rules). 

\begin{sloppypar}
Our implementation of the regret-based method is given in the ``Regret-Minimization'' estimation procedure (Algorithm \ref{alg:regret}). 
The procedure is given as input a sequence of bid profiles $(\bm{b}^t)_t$ 
(in our implementation we used $(\bm{b}^t)^{1500}_{t=751}$, excluding the first half of the game as an initial learning phase, as mentioned above), as well as 
the players' utility functions $U_i(b,\bm{b}^t_{-i}|v)$. 
Regret-Minimization estimates value $\hat{v}_i$ of each bidder $i$ as follows:
it begins by fixing the sets of possible valuations $V_i$ and of possible bids $B_i$ to consider (we set $V_i=B_i=\{1,2,...,60\}$ for all bidders). 
Then, for every possible value $v \in V_i$, it computes $Regret_i((\bm{b}^t)_t|v)$ -- the regret of player $i$ {\em had his value been} $v$ -- 
which in turn requires the computation of $Actual_i((\bm{b}^t)_t|v)$ and $Opt_i((\bm{b}^t)_t|v)$. 
Notice that in VCG, finding $Opt_i((\bm{b}^t)_t|v)$ (in line \ref{alg:regret_opt}) does not require iterating over all $b \in B_i$, as this optimal outcome is always obtained at $b=v$.
The estimate $\hat{v}_i$ is then the value that minimizes this regret.\footnote{
The specific implementation in \cite{Nekipelov2015} selected the value that minimizes the {\em relative} 
regret. 
} \footnote{
Only for a single bidder of the 120 bidders in our data, the optimal $v$ was not unique but could only be located in the range of values ($v\in [49,52]$). We used as our estimate the middle point in this range.} 
\end{sloppypar}

\begin{algorithm}
\caption{Regret-Minimization: regret-based estimation of  bidders' valuations.}
\label{alg:regret}
\begin{algorithmic}[1]
\State \textbf{Input}: actual bid sequence $(\bm{b}^t)_t$ and functions $U_i(b,\bm{b}^t_{-i}|v)$ modeling players' utilities
\State \textbf{Output}:  players' value estimates $\hat{v}_1,...,\hat{v}_n$
\Procedure{Regret-Minimization}{}
	\For{bidder $i \in N$}
		\State denote the set of possible valuations by $V_i$ and the set of possible bids by $B_i$
	\For{$v \in V_i$}
		\State $Actual_i \gets \sum_tU_i(b^t_i,\bm{b}^t_{-i}|v)$
		\State $Opt_i \gets max_{b \in B_i}\sum_tU_i(b,\bm{b}^t_{-i}|v)$ \label{alg:regret_opt}
		\State $Regret_i((\bm{b}^t)_t|v) \gets Opt_i-Actual_i$
	\EndFor
	\State $\hat{v}_i = \argmin_{v \in V_i}Regret_i((\bm{b}^t)_t|v)$
	\EndFor
\State \textbf{return} the estimates $\hat{v}_1,...,\hat{v}_n$
\EndProcedure
\end{algorithmic}
\end{algorithm}

We applied the Regret-Minimization procedure to each of the bidders in the experiment, and we evaluate the accuracy of  the estimations in the next section. 
Figure \ref{fig:regret-vs-v} presents the regret curves as a function of value, according to the types of players and the experimental conditions. The minimum point for each player serves as the estimate of his value, and is clearly visible in the graphs. As can be seen, this minimum is achieved at a higher value as we move from lower to higher types, consistent with the regret-minimization model. Also notice the asymmetry in the slopes of the regret, as well as how it changes between low and high types, 
showing that low types would have suffered low regret had their values been lower (as they tend to get the low CTR positions anyway) and higher regret had their values been higher, 
while the opposite is observed for the higher types.

\section{Evaluating Regret-Based Econometrics}\label{sec:eval}

We are now ready to evaluate the success of the Regret-Minimization method in estimating players' private valuations, in comparison with standard econometric methods that rely on the equilibrium assumption. 
The same regret-based estimation procedure (Algorithm \ref{alg:regret}) works for both the VCG and GSP auctions, with the payment rule taken into account in the calculation of the utilities, $U_i(b,\bm{b}^t_{-i}|v)$.
However, since these auction mechanisms have different equilibria, standard methods use different procedures in these two cases, and thus we perform the evaluation separately for each mechanism.

We assess the quality of the estimation methods in percentages based on the mean squares of the relative errors. Specifically, for every player $i$ whose true value is $v_i$, we compute the relative estimation error: $error_i=\frac{1}{v_i}|v_i-\hat{v}_i|$, where $\hat{v}_i$ is the value estimate for player $i$. The estimation error on a set of players $S$ is: 
$error(S)=\sqrt{\frac{1}{|S|}(\sum_{i \in S}error_i^2)}$. We base the evaluation on the second half of the auctions game: a stream of 750 auctions, excluding the first half as an initial learning phase (as mentioned above).

\subsection{Evaluating Estimations in VCG Auctions}

\begin{figure*}[t] 
\begin{subfigure}{.3\textwidth} 
  \includegraphics[scale=0.33]{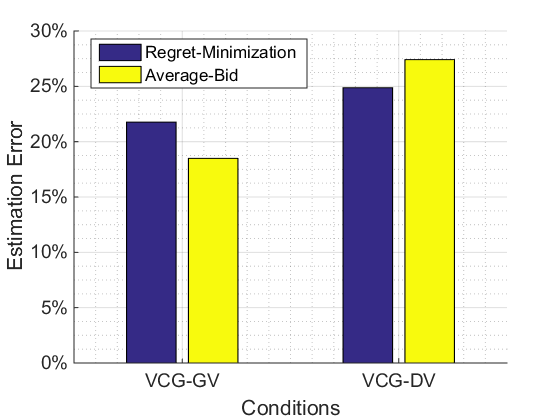}
  \caption{By conditions \label{fig:vcg-final-info}}
\end{subfigure}
\begin{subfigure}{.3\textwidth}
  \includegraphics[scale=0.33]{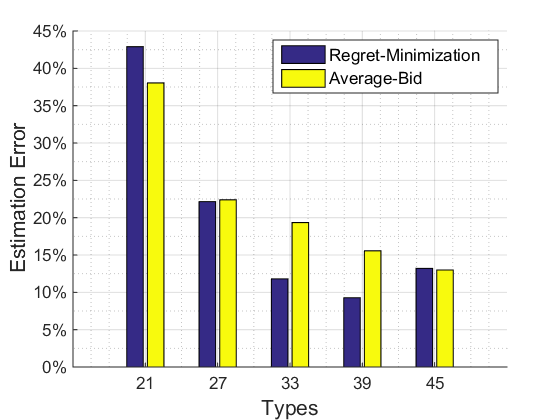}
  \caption{By player types \label{fig:vcg-final-type}}
\end{subfigure}
\begin{subfigure}{.3\textwidth}
  \includegraphics[scale=0.4]{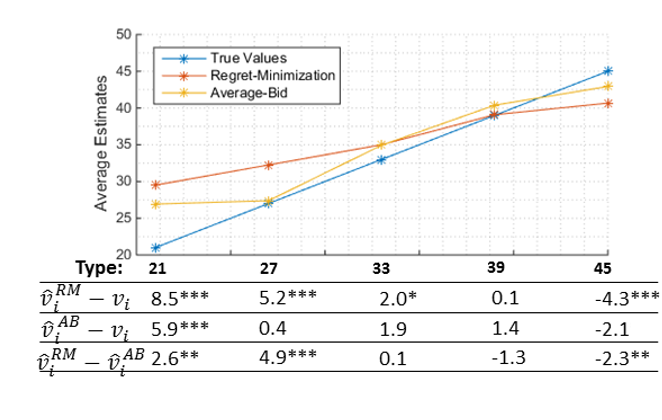}
  \caption{VCG estimates \label{fig:estimates_lines_vcg}}
\end{subfigure}
\caption{Estimation results in VCG sessions: Regret-Minimization (RM) vs. Average-Bid (AB) methods. (a) and (b) present the estimation errors by value-information conditions and by types of players, respectively. (c) presents the average estimates of the two methods according to types of players. The ``differences table'' below presents the averages of the differences between the value estimates and the true values of the bidders, and the differences between the value estimates using the two methods, by types, as indicated in the first column. 
Significant differences 
are marked by *, **, and ***, for significance at levels of 10\%, 5\%, and 1\%, respectively.
\label{fig:vcg-final}}
\end{figure*}

We start by considering the VCG auction. 
The usual econometric treatment
will note that players have dominant strategies, so it should be a strong prediction that
they all bid these dominant strategies in equilibrium.  The classic econometric method will thus
take as the model that each player bids his true value in each round, plus an error term:
$b^t_i = v_i + \epsilon^t_i$, where $v_i$ is the true value and $b^t_i$ is the bid in round $t$.  
The estimate of
the hidden value $v_i$ from the visible data of the $b^t_i$'s that were played in a sequence of $T$ auctions will be 
$\hat{v}_i = (\sum_{t=1}^T b^t_i)/T$, as this average minimizes the sum of the squares of
 $\epsilon^t_i$: $\sum_t (v - b^t_i)^2$.   
We applied the Regret-Minimization method and the standard method of taking the average bid (``Average-Bid'') 
to estimate the valuations of the bidders in the VCG sessions. 
Overall, considering all 60 VCG bidders, Regret-Minimization and Average-Bid performed very similarly, with total estimation errors of 23.37\% and 23.38\%, respectively. The errors distributed similarly, both with high variance between players: $\mu=17.69\%$ and $\sigma=15.39\%$ using Regret-Minimization, and $\mu=17.37\%$ and $\sigma=15.79\%$ using Average-Bid. The similarity persists when considering errors by information settings or by types of players, as can be seen in Figures \ref{fig:vcg-final-info} and \ref{fig:vcg-final-type}. 
Thus, the estimation obtained by the Regret-Minimization method, which is general for all auction formats, does not fall from the estimation of the standard approach, even though the specific equilibrium prediction for VCG is strong and is of the simple strategy of bidding the true value.

Figure \ref{fig:vcg-final-info} shows that the estimation errors of both methods are somewhat larger in the deduced-value setting (DV) than in the given-value setting (GV), as could be expected, and the regret-based method seems somewhat better in the former setting and somewhat worse in the latter. 
However, these effects were not statistically significant.

A significant effect is revealed for both methods when considering errors by types of players: as expected from the finding in Section \ref{sec:regret} that the low types deviate significantly from playing rationally, 
the estimation errors of both methods for the low types are far larger than those for the high types (see Figure \ref{fig:vcg-final-type}). 
Specifically, for each of the two methods, there is a significant negative correlation between the estimation error and the player type ($N=60$, $\rho=-0.65$ and $\rho=-0.52$ for Regret-Minimization and Average-Bid, respectively, $p<0.001$).
The total estimation error of both methods for the lowest-type players is reaching around 40\%, while for the highest-type players the error is much lower -- around 13\%.
Despite the overall similarity between the two methods, Figure \ref{fig:vcg-final-type} shows that while Regret-Minimization is more accurate for the middle-valued types, Average-Bid is more accurate for the lower-type players.\footnote{
This pattern is robust for different choices that we tried of initial learning phases (i.e., other than the first half of the game), and for some choices the interaction between method and player types reaches statistical significance at 
5\% level.} 
In addition, as opposed to the significant and gradual differences between estimation errors for different types using Regret-Minimization (consistent with the gradual pattern of regret levels shown in Figure \ref{fig:total_regret_type}),  using Average-Bid the differences are only due to the error in estimating the lowest type's valuation, which was significantly higher than that of each of the other types. 
But still, the errors of the two methods were not significantly different at the 5\% level also when tested for each type separately.

Finally, we found that Regret-Minimization and Average-Bid are different in terms of the average estimates that they give. Figure \ref{fig:estimates_lines_vcg} compares the average estimates obtained by the two methods and the bidders' true values (by types). 
As can be seen, Regret-Minimization significantly overestimates the valuations of the three lower-type bidders, and significantly underestimates the valuations of the highest-type bidders. The direction of mistakes by Regret-Minimization gradually changes from low to high types. These findings suggest that the low-type bidders tend to play as if their valuation were higher and the opposite holds for the bidders with the highest value, who tend to play as if their valuation were lower than it really was. Interestingly, these tendencies do not clearly arise from the average estimates of the Average-Bid method, and thus must be hidden in the dynamic strategies played by the players relative to their opponents. 
In Section \ref{sec:possible_improvements} we demonstrate how we may improve the accuracy of estimations by taking advantage of the differences between the two methods.

\subsection{Evaluating Estimations in GSP Auctions}

For the GSP auction the situation is much more complicated for equilibrium-based econometric methods since
there are no dominant strategies and there exist multiple equilibria.  There are two basic 
approaches in the literature for deducing bidders' valuations in GSP auctions. 
In \cite{Varian2007} it is suggested that the players should reach the
equilibrium of the full-information one-shot GSP game that gives the VCG-prices (hence the ``VCG-like'' equilibrium).  Assuming that this is indeed the case,
then at each time step $t$ in a sequence of auctions, one may deduce values $\hat{v}_i^t$ for all players $i$ from the actual bids $b_i^t$, such that the bids are this VCG-like equilibrium of these deduced values.
The final estimate is then the average of these $\hat{v}_i^t$.  
Some complications arise when this is attempted on real data since it is often the case that the bids 
do not correspond to an equilibrium of any tuple of values. 
In these cases we follow \cite{Varian2007} and perturb the bid observations in the minimal possible way so as to satisfy the equilibrium constraints, and set the final estimates to the perturbed values.\footnote{
In fact, only 13.3\% of the auctions were consistent with the equilibrium inequalities without perturbing their data. However, similar to \cite{Varian2007}, we observed that the required perturbations were relatively small.} 
These and other complications of the ``VCG-like-NE'' method
are discussed in Appendix \ref{app:hal} and in \cite{Varian2007}. 

A 
second 
method was suggested by \cite{Athey2010} where bidders participate in a large number of auctions, and receive feedback that can vary from auction to auction. 
The basic assumption is that each bidder is best-responding to the {\em distribution} that he faces (by placing a single bid).
Specifically, given a sequence of auctions, define functions $Q_i(b_i)$ and $TE_i(b_i)$ as the expected CTR and the expected total expenditure, respectively, of bidder $i$ by bidding $b_i$. 
Thus, his expected utility with valuation $v$ is $Q_i(b_i) \cdot v - TE_i(b_i)$.
Against {\em smooth} distributions the best bid would be a strictly increasing function of the value. 
In these cases, the valuation of bidder $i$ who maximizes his expected utility by bidding $b_i$ can be recovered using the first-order condition by $\hat{v}_i = \frac{\partial TE_i(b_i)/\partial b_i}{\partial Q_i(b_i)/\partial b_i}$.
When applying this method to actual data complications arise, and there are many possible implementations. 
We have tested several implementation variants,  
and in the implementation we chose (referred to as the ``Best-Response'' method), 
we used the average bid that a bidder played as his best-response to the distribution of the bids of the others (since bids were not constant), 
and found the value by optimizing directly using grid search (since the empirical derivatives had their own complications). 
Details of implementation and complications of this method are discussed in Appendix \ref{app:susan} and in \cite{Athey2010}.

\begin{figure*}[t] 
\begin{subfigure}{.3\textwidth}
  \includegraphics[scale=0.33]{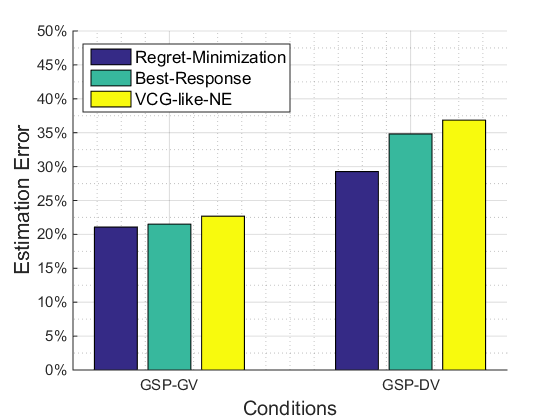}
  \caption{By conditions}
  \label{fig:gsp-final-info}
\end{subfigure}
\begin{subfigure}{.3\textwidth} 
  \includegraphics[scale=0.33]{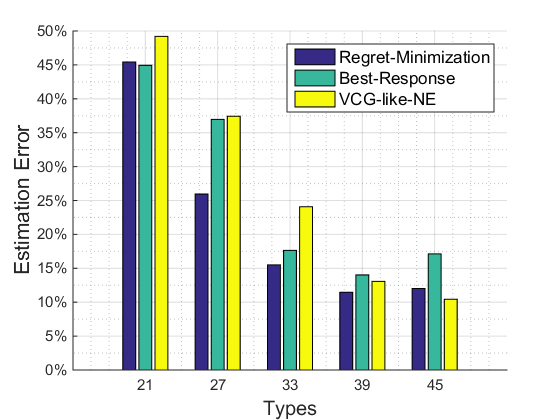}
  \caption{By player types}
  \label{fig:gsp-final-type}
\end{subfigure}%
\begin{subfigure}{.3\textwidth}
  \includegraphics[scale=0.5]{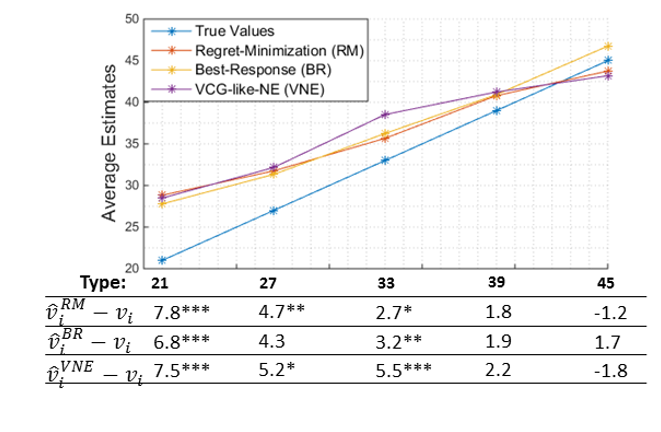}
  \caption{GSP estimates \label{fig:estimates_lines_gsp}}
\end{subfigure}
\caption{Estimation results in GSP sessions: Regret-Minimization (RM) vs. Best-Response (BR) vs. VCG-like-NE (VNE) methods. (a) and (b) present the estimation errors by value-information conditions and by types of players, respectively. (c) presents the average estimates of the three methods according to types of players. The ``differences table'' below presents the averages of the differences between the value estimates and the true values of the bidders, by types, as indicated in the first column. 
Significant differences 
are marked by *, **, and ***, for significance at levels of 10\%, 5\%, and 1\%, respectively.
\label{fig:gsp-final}
}
\end{figure*}

Figures \ref{fig:gsp-final-info} and \ref{fig:gsp-final-type} compare the estimation error 
of the regret-based method with those of the two ``classic'' equilibrium-based methods. 
Overall, 
the regret-based method 
tends 
to be 
better
than the other two methods, but 
the difference is statistically significant at the 5\% level only in a comparison with the VCG-like-NE method.\footnote{The error using the Best-Response method was lower than the error using VCG-like-NE at a significance level of 10\%.}
Specifically, over all 60 GSP bidders, the total estimation errors obtained using Regret-Minimization, Best-Response, and VCG-like-NE were 25.51\%, 28.94\%, and 30.60\%, respectively. 
The average of the estimation errors 
using Regret-Minimization, Best-Response and VCG-like-NE, 
were 18.07\%, 20.04\%, and 21.45\%, respectively, and the standard deviations were 18.00\%, 20.87\% and 21.82\%, respectively.

Considering estimation errors by information settings (Figure \ref{fig:gsp-final-info}), 
the errors  of all three methods are somewhat larger in the deduced-value setting (DV), in which case the regret-based method has an advantage over the other two methods. While this appears to be quite consistent (for different selections of the initial learning phase), it is only partially statistically significant.

As in the case of VCG, the significant effect is revealed when considering errors by player types (see Figure \ref{fig:gsp-final-type}): 
all three methods succeed better in estimating valuations of the higher-type players (who succeeded better in minimizing their regret) than of the lower-type players (who remained with high levels of regret, see Section \ref{sec:regret}), for whom all methods perform very poorly.
Specifically, for each of the three methods, the estimation errors are negatively correlated with the player type ($N=60$, $\rho=-0.59$, $\rho=-0.47$, and $\rho=-0.57$ for Regret-Minimization, Best-Response, and VCG-like-NE, respectively, $p<0.001$).
When comparing estimation errors for each type separately, 
the differences are statistically significant only for types 27 and 33: the error using Regret-Minimization is lower at 5\% level than the error using VCG-like-NE for types 27 and 33, and Best-Response is also better than VCG-like-NE at 5\% level for type 33.

Finally, we found that the three methods are similar in terms of the average estimates that they give. As can be seen in Figure \ref{fig:estimates_lines_gsp}, all three methods tend to overestimate the valuation of the three lowest types. 
Thus, also in GSP, where bidders are expected to shade their bids, lower types play as if their value were higher, in consistency with their high levels of regret described in Section \ref{sec:regret}. 
The estimations of the two highest types are not significantly in a particular direction relative to the 
values. The direction of mistakes changes gradually across types, as in the case of VCG.

\section{Possible Improvements}\label{sec:possible_improvements}
We have seen, in the previous section, that the regret-based estimation method gives value estimates that are competitive with standard econometric methods both in GSP and in VCG. Thus, the weaker assumptions that the regret-based method makes seem sufficient for this estimation task. 
However, we have also seen how {\em all} methods have quite high errors, particularly on the lower-type players, who, as we found, tend not to follow the methods' underlying assumption of rationality. 
Therefore we ask: how can we improve the accuracy of the estimations using regret-based econometrics and our understanding of the biases of human players?\footnote{One general approach one may try is to ``clean'' the data, i.e., remove outlier observations that might bias the estimates. However, our attempts in this direction did not improve the estimation accuracy and even slightly increased the error, indicating that this error is not a result of the ends of the bid distributions but is inherent in the bidders' behavior.} 

First, it may be useful to combine estimates of the different estimation methods, taking advantage of the differences between them, 
as the different methods catch different aspects of behavior and may have different sources of mistakes. 
For example, we averaged the estimates of the Regret-Minimization and Average-Bid methods for each of our VCG bidders, and the estimation error of the combined method was lower than the error of each of the two methods separately (23.37\%, 23.38\%, and 21.73\%, for the Regret-Minimization, Average-Bid, and the combined method, respectively; however the differences in our results were 
statistically significant at the 5\% level only with the Regret-Minimization method).

\begin{figure}
\begin{subfigure}{.48\linewidth}
  \includegraphics[scale=0.5]{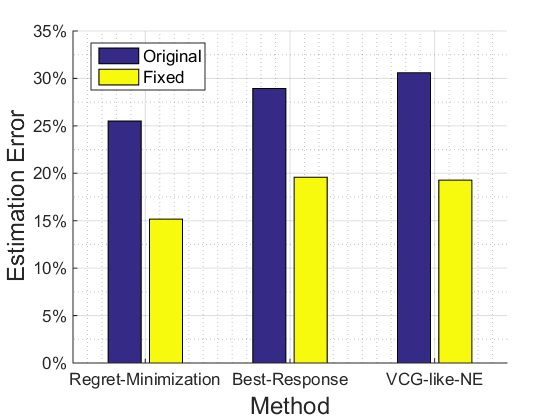}
  \caption{Improving by ``un-biasing'' the estimates according to player types. Original and fixed (un-biased) estimation errors for the GSP bidders.} 
  \label{fig:fix_bias_GSP_allMethods_total}
\end{subfigure}
\hspace{0.2cm}
\begin{subfigure}{.48\linewidth} 
  \includegraphics[scale=0.5]{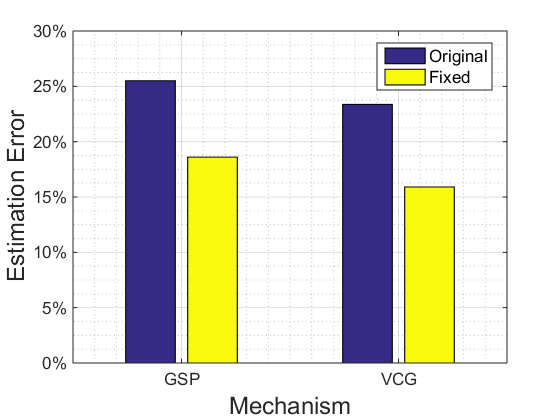}
  \caption{Improving by the ``regret-based average'' method, which estimates using the entire regret curve rather than just the minimal regret point.}
  \label{fig:improve_avg}
\end{subfigure}%
\caption{Possible improvements of the value estimates.}
\label{fig:gsp-final}
\end{figure}

A second approach is to ``un-bias'' the estimates according to player types, taking advantage of our finding that 
the bias of all methods 
was correlated with the types of players.
Since this seems to result from systematic differences in behavior between player types, where the lower-ranked players tend (irrationally) to play as if their value were higher, it is useful to know 
to what extent we could have improved the estimates had we known how to quantify and correct this bias.
We demonstrate the potential gains of un-biasing the estimates on our GSP bidders: 
for every bidder, we calculated his most frequent position in the auction, 
and used this information to ``un-bias'' our estimate according to the average 
bias of the estimates of his (deduced) rank.
For example, let $\hat{v}$ be the estimate using Regret-Minimization of a GSP bidder who was most frequently observed in the last position. 
Then, we use the finding that on average Regret-Minimization overestimates the lowest type as 137\% of his true value, and we fix the estimate accordingly, so that $\hat{v}_{fixed}=\hat{v} / 1.37$. 
This correction of the bias significantly improved the results of all three methods for the GSP bidders, as can be seen in Figure \ref{fig:fix_bias_GSP_allMethods_total}.
Of course, since this bias-correction factor (137\% in the example) came from within the sample using the real values, one can only interpret these reduced estimation errors as an upper bound on the gains from correcting by types. Our demonstration here suggests that these gains may be substantial, indicating that it is worthwhile to further understand the behavioral tendencies of the different types. 


Finally, we suggest improving the estimates of the Regret-Minimization method by taking advantage of the shape of the regret curves, which we found to be different for the different types of players (see Figure \ref{fig:regret-vs-v}). Since we observed that human bidders do tend to minimize their regret, but do not minimize perfectly, we suggest setting the estimate based on the entire regret curve rather than just on the minimal regret point. For example, we set the value estimate of a bidder $i$ to the weighted average of all possible valuations, with weights that are proportionally decreasing in the regret: 
$\hat{v}_i = \frac{\sum_{v\in V_i}v \cdot Regret_i^{-1}(\cdot |v)}{\sum_{v\in V_i}Regret_i^{-1}(\cdot |v)}$. 
This ``regret-based average'' method is completely applicable based on observed data only, and, as can be seen in Figure \ref{fig:improve_avg}, significantly improves the accuracy of the estimates for our bidders. In a separate paper we explore this approach further \cite{quantalregret}.

\section{Conclusions and Further Directions}

We have demonstrated that, at least in our repeated ad-auction experiment, players do act to reduce 
their regret, 
and that the 
regret-based method suggested by \cite{Nekipelov2015} for estimating players' values from their bids is (at least) competitive with 
``classic'' equilibrium-based econometric methods.  
We find this to be especially significant due to the generality and simplicity of the regret-based method: while ``classic'' econometric methods
require choosing between different 
interpretations of equilibria 
as well as among many significant implementation details, and to tailor the 
method to the specific equilibrium assumptions, the regret-based econometric method 
does not require any specific tailoring and hence is much easier to specify and implement.
We furthermore speculate that this simplicity may reasonably allow this method to be further improved, 
and we suggested 
possible improvement approaches of this form. 

En route we also observed irrational bidding in our experiment and identified a distinct human bias: players with ``low'' types tend to overbid
leading to high regret, and resulting in increased estimation errors in all the econometric methods that we tried. 
It is an interesting challenge 
to use our understanding of this behavioral bias to improve the estimates, and we demonstrated that the improvements based on this bias may be substantial. 
Regret-minimization is a simple and powerful method that can be easily modified to capture different behavioral phenomena and to take advantage of insights from behavioral disciplines. 
Obviously, further evaluation of regret-based econometrics in more scenarios is called for.

%
%

%
\bibliographystyle{apalike} 
\bibliography{regret-bib}  
%
%
\appendix

\section*{APPENDICES}

\section{Summary of Existing Estimation Methods for GSP}
The case of GSP auctions is more complicated for standard econometric methods that are based on the equilibrium assumption, since it is less clear to which equilibrium bids should converge; unlike VCG, in GSP players do not have dominant strategies and there might be multiple equilibria which might be complex in the repeated game. 
There are two main approaches in the literature for deducing bidders' valuations in GSP auctions: the first assumes bidders reach the equilibrium of the full-information game, that gives the VCG-prices (thus ``VCG-like'' equilibrium) \cite{Varian2007, EOS2007}; the second refers to the uncertainty in the game and assumes bidders best respond to the distributions they experience \cite{Athey2010}.

\subsection{The VCG-like-NE Method} \label{app:hal}

\subsubsection{Method Overview}
\cite{Varian2007} and \cite{EOS2007} suggested it is reasonable to assume that bids in GSP implement equilibria of the induced full-information one-shot game. 
They focus on the set of ``Symmetric Nash Equilibrium''
\footnote{\cite{Varian2007} presented the ``Symmetric Nash Equilibria'', which are Nash equilibria in which no bidder can benefit from exchanging position and payment with any of the other bidders. \cite{EOS2007} presented (almost) equivalent class of equilibria they termed ``Locally Envy-Free Equilibria''.
}
(SNE), and particularly point to the specific SNE that gives the VCG-equilibrium prices, as the equilibrium that is most plausible to be implemented by the players. 
\cite{Varian2007} suggested a procedure for deducing bounds on valuations assuming bids in every auction are at an SNE, and applied it to actual ad auctions. 
Here we follow this procedure as the first ``classical'' equilibrium-based estimation of the bidders' values, that is compared to the Regret-Minimization procedure.

Assuming a given bid profile is an SNE, \cite{Varian2007} showed that the bidders' (unobserved) values are bounded by the (observed) incremental cost per click (ICC) of moving up or down one position. 
Specifically, if $(b_1,...,b_n)$ is an SNE (and bids are numbered by decreasing order), then the value $v_k$ of the bidder in the $k$'th ($k>1$) position is bounded by

\begin{equation} \label{eq:hal_bounds}
\frac{\alpha_{k-1}b_k - \alpha_k b_{k+1}}{\alpha_{k-1}-\alpha_k} \geq v_k \geq \frac{\alpha_kb_{k+1} - \alpha_{k+1} b_{k+2}}{\alpha_k-\alpha_{k+1}}
\end{equation}

For the highest ranked bidder, this equilibrium assumption does not provide an upper bound, and implies only that his value must be at least as high as the value of the second highest bidder.
If the bids are at the VCG-like equilibrium, where prices are minimal among SNE prices, then values equal the upper bound in the inequality, i.e., each value equals the ICC of moving to the next higher position. 
Applying (\ref{eq:hal_bounds}) iteratively in an auction of five players as in our setting, 
gives that the valuations corresponding to a VCG-like equilibrium bids $(b_1,...,b_5)$ (in decreasing order) must satisfy the following

\begin{equation} \label{eq:hal_values}
v_1 \geq \frac{\alpha_1b_2 - \alpha_2b_3}{\alpha_1-\alpha_2} = \\
v_2 \geq \frac{\alpha_2b_3 - \alpha_3b_4}{\alpha_2-\alpha_3} = \\ 
... \geq \frac{\alpha_4b_5}{\alpha_4-\alpha_5} =
v_5 \geq b_5
\end{equation}

When using the ICC to estimate valuations of each of the bidders separately on real data, the estimates do not necessarily satisfy the SNE inequalities in (\ref{eq:hal_values}). 
\cite{Varian2007} viewed these inconsistencies as resulting from the gap between the full information case and the uncertainty in real auctions and suggested to resolve them by perturbing 
the  
observations, in the ``minimal'' possible way as to satisfy the equilibrium constraints, and set the final estimates to the perturbed values.

\subsubsection{Our Implementation}

Algorithm \ref{alg:vcg-like} specifies the ``VCG-like-NE'' estimation procedure, which we applied to each of the GSP experiment sessions, 
as the first comparison alternative to the regret-based estimation method. As in Regret-Minimization (see Section \ref{sec:method}), 
we focused on the second half of the auctions game, excluding the first half as the initial learning phase (i.e., using $(\bm{b}^t)^{1500}_{t=751}$).

\begin{algorithm}
\caption{VCG-like-NE: estimating bidders' valuations in GSP sessions, assuming bids at every auction are at the Nash equilibrium that gives the VCG prices.}
\label{alg:vcg-like}
\begin{algorithmic}[1]
\State \textbf{Input}: actual bid sequence $(\bm{b}^t)_t$ and parameters of the GSP auction $\bm{\alpha}$.
\State \textbf{Output}: players' value estimates $\hat{v}_1,...,\hat{v}_n$.
\Procedure{VCG-like-NE}{}
	\For{\textbf{each} $1 \leq t \leq T$}
	\State order the bidders by decreasing order of $\bm{b}_i^t$
	\State $\hat{v}^t_1 \gets Infinity$ 
	\State $\textbf{for all}$ $1 < i \leq n$: $\hat{v}^t_i \gets \frac{\alpha_{i-1}b^t_i - \alpha_i b^t_{i+1}}{\alpha_{i-1}-\alpha_{i}}$ \Comment{set ICC as	in (\ref{eq:hal_values}) }
		\If{$\hat{\bm{v}}^t$ is not consistent with (\ref{eq:hal_values})} 	\Comment{fix inconsistencies} \label{alg:perturb_start}
			\State $\bm{d} \gets GetMinimalPerturbations(\bm{b}^t)$
				\State $\textbf{for all}$ $1 < i < n$:  $\hat{v}^t_i \gets \frac{\alpha_{i-1}b^t_i - \alpha_ib^t_{i+1}d_i}{\alpha_{i-1}-\alpha_{i}}$ \label{alg:set_perturb}
				\State $\hat{v}^t_n \gets \frac{\alpha_{n-1}b^t_n}{\alpha_{n-1}-\alpha_n}$ 
		\EndIf \label{alg:perturb_end}
		\State $\hat{v}^t_1 \gets \hat{v}^t_2$ 	\Comment{set highest value} \label{alg:set_highest_value}
	\EndFor
	\State $(\hat{v}_1,...,\hat{v}_n) \gets (\frac{1}{T}\sum_t(\hat{v}^t_1),...,\frac{1}{T}\sum_t(\hat{v}^t_n))$ 
	\State \textbf{return} the estimates $\hat{v}_1,...,\hat{v}_n$
\EndProcedure
\\
\Function{GetMinimalPerturbations}{$\bm{b}^t$}
    \State \Return \begin{varwidth}[t]{\linewidth}
      $\argmin_{\bm{d}} \sum_i(d_i-1)^2 $ $s.t.$ for all $i$:\par
        \hskip\algorithmicindent $\frac{\alpha_{i-1}b_i - \alpha_ib_{i+1}d_i}{\alpha_{i-1}-\alpha_i} \geq \frac{\alpha_ib_{i+1} - \alpha_{i+1}b_{i+2}d_{i+1}}{\alpha_i-\alpha_{i+1}}$ 
      \end{varwidth}
\EndFunction
\end{algorithmic}
\end{algorithm}

\begin{figure}
\centerline{\includegraphics[scale=0.5]{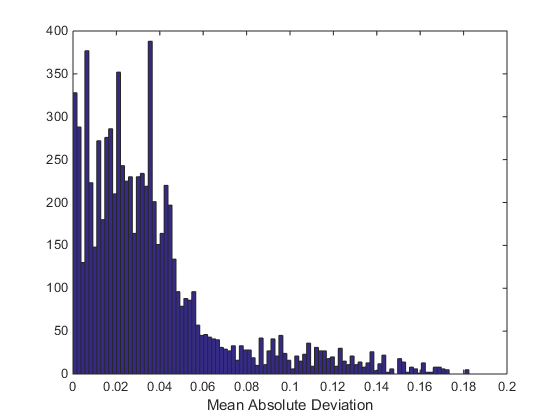}}
\caption{The distribution of the mean absolute deviations from an SNE of GSP auctions in the experiment (the second half of the auctions game). 
\label{fig:perturbations_hist}}
\end{figure}

\newpage

\subsubsection{Implementation Remarks}

\paragraph{Perturbations}

When the observed bids $\bm{b}^t$ were not consistent with the SNE model (lines \ref{alg:perturb_start}-\ref{alg:perturb_end}), we fixed the inconsistencies (i.e., satisfied the equilibrium inequalities in (\ref{eq:hal_values})), as in \cite{Varian2007}, by perturbing for every bidder $i\in \{2,3,4\}$ the bid that follows him in the ranking of bidders $b_{i+1}$, 
which is the only variable in these inequalities that each bidder could not directly observe. 
As in \cite{Varian2007}, we chose the minimal perturbations by solving the quadratic programming problem in 
subroutine \textit{GetMinimalPerturbations}, and set the estimates to the perturbed valuations.

\begin{figure}
\begin{subfigure}{.5\textwidth} 
  \includegraphics[scale=0.53]{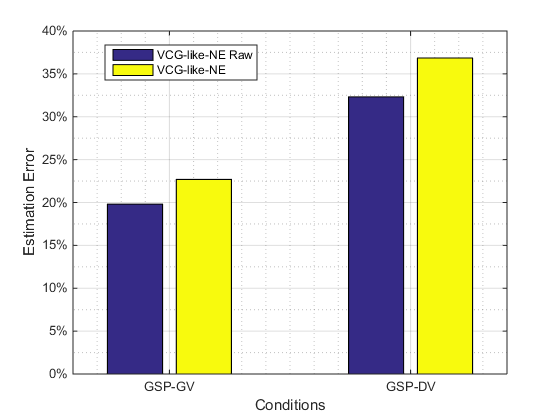}
  \caption{By value-information conditions}
\end{subfigure}
\begin{subfigure}{.5\textwidth}
  \includegraphics[scale=0.53]{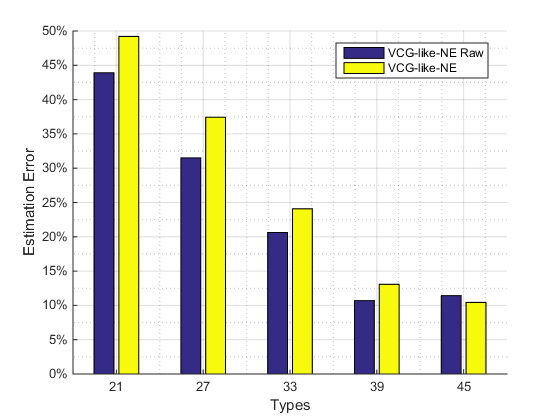}
  \caption{By player types}
\end{subfigure}
\caption{Estimation errors in GSP sessions using VCG-like-NE: errors of estimates using perturbations (VCG-like-NE) vs. using the raw ICC estimates without perturbing the data (VCG-like-NE Raw). \label{fig:pert_vs_raw}}
\end{figure}

Only 13.3\% of the auctions were consistent with the equilibrium inequalities without perturbing their data. However, similar to \cite{Varian2007}, we observed that the required perturbations were relatively small. Figure \ref{fig:perturbations_hist} shows the distribution of the mean absolute deviation from an SNE in each auction, defined as 
$\frac{1}{5}\sum^5_{i=1}|d_i-1|$ (setting $d_1=d_5=1$). The histogram looks similar to the histogram in \cite{Varian2007}, with an average of 3.65\% and median 2.88\%. 
Interestingly, perturbing the data did not improve estimation results, but, in fact, slightly increased the estimation errors, as can be seen in Figure \ref{fig:pert_vs_raw}. 
Over all 60 GSP bidders, the estimation error was 30.60\% when using perturbations and 26.80\% using the raw ICC without perturbing the data (the difference was not statistically significant).

\paragraph{Highest Ranked Bidder}

The VCG-like-NE method provides only a lower bound for the highest ranked bidder and thus it is unclear what should be the estimation of his value. We tried two reasonable implementations: estimating his value as the second highest value, or as the maximum between the second highest value and his own bid in the auction. We chose the first implementation (in line \ref{alg:set_highest_value}) since it gave a slightly lower estimation error (30.60\% vs. 30.76\%, respectively, over all GSP bidders).

\subsection{The Best-Response Method} \label{app:susan}

\subsubsection{Method Overview}
\cite{Athey2010} suggested a method for deducing bidders' valuations in GSP auctions that takes into account the uncertainty in the game. They consider a framework in which bidders participate in a large number of auctions, and receive feedback that can vary from auction to auction. 
The basic assumption is that bidders are best responding to the distributions that they experience. 

Given a sequence of auctions, define functions $Q_i(b_i)$ and $TE_i(b_i)$ as the expected CTR and the expected total expenditure, 
respectively, of bidder $i$ who bids $b_i$ through all these auctions, with the expectation taken over the 
distribution of the uncertainty.
Thus, the expected utility of bidder $i$ who bids $b_i$ and has value per click $v$ is $Q_i(b_i) \cdot v - TE_i(b_i)$.
With sufficient uncertainty in the environment, these functions are strictly increasing and differentiable in $b_i$. 
In these cases\footnote{When this is not the case, \cite{Athey2010} show how to derive bounds on valuations. 
}, the valuation of bidder $i$ who maximizes his expected utility by bidding $b_i$ in the auctions' sequence, can be recovered using the first-order condition (FOC), as follows 

\begin{equation} \label{eq:susan_foc}
v = \frac{\partial TE_i(b_i)/\partial b_i}{\partial Q_i(b_i)/\partial b_i}
\end{equation} 

Note that in order for $b_i$ to be a best-response bid for a bidder with value $v$, it is also required to verify for global optimality of the bid, as (\ref{eq:susan_foc}) guarantees only local optimality.

The original model considers that bidders face uncertainty about ``quality scores''\footnote{
In actual sponsored search auctions, ads are ranked not only by their bids but also by the quality score that the search engine assigns to each ad, a score that can vary between auctions. 
} 
and the set of competitors.  These specific sources of uncertainty do not exist in our controlled experiment, but the uncertainty in our case is due to the distribution of bids of the other bidders.

\subsubsection{Our Implementation}

Algorithm \ref{alg:best-response} specifies the ``Best-Response'' estimation procedure, which is our implementation of the method suggested by \cite{Athey2010}, and is 
the second ``classical'' equilibrium-based method that we compare with the regret-based estimation.
Similar to the Regret-Minimization procedure (see Section \ref{sec:method}), we focus on the second half of the game and use $(\bm{b}^t)^{1500}_{t=751}$.

\begin{algorithm}
\caption{Best-Response: estimating bidders' valuations in GSP sessions, assuming bidders best-respond to the distributions they experience. }
\label{alg:best-response}
\begin{algorithmic}[1]
\State \textbf{Input}: actual bid sequence $(\bm{b}^t)_t$ and parameters of the GSP auction $\bm{\alpha}$.
\State \textbf{Output}: players' value estimates $\hat{v}_1,...,\hat{v}_n$.
\Procedure{Best-Response}{}
	\For{bidder $i \in N$}
		\State denote the set of possible valuations by $V_i$ and of possible bids by $B_i$
		\For{$b \in B_i$}
			\State $Q_i(b) = \frac{1}{T}\sum_t\alpha_{position_i (b, \bm{b}^t_{-i})}$  
			\State $TE_i(b) = \frac{1}{T}\sum_t payment_i (b, \bm{b}^t_{-i})$					
		\EndFor
		\For{$v \in V_i$}
			\State $BR_i(v) = \argmax_{b \in B_i} (Q_i(b)\cdot v - TE_i(b))$
		\EndFor
		\State $b^*_i = \frac{1}{T}\sum_t b^t_i$
		\State $\hat{v}_i = v$ s.t., $b^*_i \in BR_i(v)$ \label{alg:susan_estimate}
	\EndFor
\State \textbf{return} the estimates $\hat{v}_1,...,\hat{v}_n$
\EndProcedure
\end{algorithmic}
\end{algorithm}

\subsubsection{Implementation Remarks} \label{subsec:susan_remarks}

The ``Best-Response'' procedure estimates valuations separately for each bidder, assuming the bidder's average bid $b^*_i$ in the given sequence of auctions is a best-response to the distribution of the bids of the others in these auctions. Rather than using the empirical derivatives, as in \cite{Athey2010}, we optimized directly using a grid of $V_i=B_i=\{1,2,...,60\}$ for all bidders $i \in N$ (see discussion below). 
Specifically, for each value $v \in V_i$, we compute the set of fixed best-responses $BR_i(v)=\argmax_{b\in B_i}(Q_i(b) \cdot v-TE_i(b))$. The estimate for bidder $i$ is set to the value for which the 
average bid belongs to its best-response set (in line \ref{alg:susan_estimate}).  
If there is no such a value (i.e., $b^*_i$ is not a best-response to any of the integer valuations on the grid), then the estimate is set to the average of the two adjacent values $v$ and $v+1$ for which $b^*_i > \max(BR_i(v))$ and $b^*_i < \min(BR_i(v+1))$.\footnote{For two of the bidders the average bid $b^*_i$ was larger than the maximal best-response to the maximal value on grid. In these cases we set the bidders' estimated values to the maximal value on the grid. In two other cases, where $b^*_i$ was a best response to a range of valuations rather than to a unique value, we set the estimation to the middle point in the obtained range and compute the related estimation error relative to the range.
}

\paragraph{Implementation Variants}

\begin{figure}
\centerline{\includegraphics[scale=0.65]{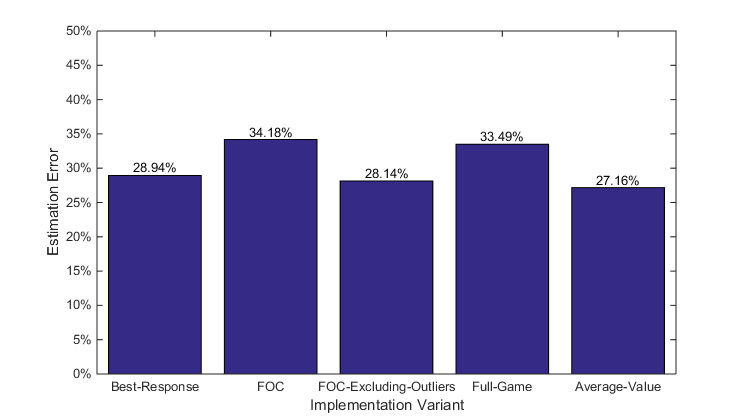}}
\caption{Estimation errors of the Best-Response variants over all GSP bidders. The variants are described in Section \ref{subsec:susan_remarks}. \label{fig:susan_variants}}
\end{figure}

One can think of many other implementations to the method suggested by \cite{Athey2010}. Figure \ref{fig:susan_variants} presents the estimation errors obtained using several of the variants that we have tested, each variant changes the Best-Response procedure in one aspect. 
The first ``FOC'' variant that we looked at follows \cite{Athey2010} directly and finds the value using the first-order condition in equation (\ref{eq:susan_foc}) rather than using direct search on a grid. 
This was sensitive to outlier bids and achieved worse estimation results.   The ``FOC-Excluding-Outliers'' presents the estimation errors obtained when modifying the FOC
method by removing outliers: computing the 
average $b^*_i$ excluding bids $b^t_i$ that were more than two standard deviations away from the average $\frac{1}{750}\sum^{1500}_{t=751}b^t_i$. 
Two other variants that we attempted are: 
(1) the ``Full-Game'' variant considers the entire game rather than focusing on the second half of the game, and specifically computes for every bidder the average 
bid $b^*_i$ and the functions $Q_i$ and $TE_i$ in respect to the 1500 auctions of the game; 
(2) the ``Average-Value'' variant computes an estimate $\hat{v}^t_i$ separately for every auction $t$ (specifically by using $b^t_i$ rather than $b^*_i$, and executing the procedure for every auction $t$ to obtain $\hat{v}^t_i$), and determines the final estimate $\hat{v}_i$ as the average of these values. This is a different approach for handling the difference from the situation in \cite{Athey2010}
where bids were assumed to be constant. 
As can be seen in Figure \ref{fig:susan_variants}, these variants were either worse or just marginally better than our choice of the basic ``Best-Response'' method. The variability does highlight,
however, the breadth and significance of implementation details when using this method.

\end{document}